\documentstyle[12pt]{article}
\newcommand{\svec}[1]{ \stackrel{\rightarrow}{#1} }

\newcommand{\define}{ \stackrel{\triangle}{=} }

\def\be{\begin{equation}}
\def\ee{\end{equation}}
\def\ba{\begin{array}}
\def\ea{\end{array}}

\begin{document}
\title{\bf Determination of Cosmological Constant from
		Gauge Theory of Gravity }
\author{{Ning Wu}$^1$
\thanks{email address: wuning@ihep.ac.cn},
Germano Resconi$^2$, 
Zhan Xu$^3$, Zhi-Peng Zheng$^1$, \\
Da-Hua Zhang$^1$ ,Tu-Nan Ruan$^4$  
\\
\\
{\small Institute of High Energy Physics, P.O.Box 918-1,
Beijing 100039, P.R.China}$^1$  \\
\\
{\small Mathematical and Physical department, Universita 
Cattolica del Cacro Cuore,}  \\
{\small Via Trieste 17 , Brescia Italy}$^2$  \\
\\
{\small Institute for Advanced Study,
Tsinghua University, Beijing 100080, P.R.China}$^3$\\
\\
{\small Dep. of Modern Phys., University of Science and Technology
of China,}\\ 
{\small Hefei, Anhui 230026, P.R.China}$^4$
}
\maketitle
\vskip 0.2in

~~

\vskip 0.3in

\begin{abstract}

Combining general relativity and gravitational gauge theory,
the cosmological constant is determined theoretically. 
The cosmological constant is related to the average 
vacuum energy of gravitational gauge field.
Because the vacuum energy of gravitational gauge field
is negative, the cosmological constant is positive, which
generates repulsive force on stars to make the expansion
rate of the Universe accelerated.
A rough estimation of it gives out its magnitude of the 
order of $10^{-52} m^{-2}$, which is well constant with 
experimental results.

\end{abstract}
\newpage

\Roman{section}

\section{Introduction}

Cosmological constant is first introduced into physics
by A.Einstein in 1917\cite{1}. In that paper, he pointed
that a small and positive cosmological constant is needed. 
At present, it is known that the cosmological constant
has evoked much controversy in both cosmology and 
particle physics\cite{2,3}. Recent observations of 
high-redshift supernovae seem to suggest that the global 
geometry of the Universe may be affected by a small
positive cosmological constant, which acts to accelerate
the expansion rate with time\cite{4,5,6,7,8}.\\

Some theoretical models are proposed to explain the
cosmological constant problems\cite{9,10}. In this
paper, we will use another more fundamental way to
determine the cosmological constant. Our analysis
based on a new quantum theory of gravity --- QUANTUM
GAUGE THEORY OF GRAVITY\cite{11,111}, which is formulated
completely in the framework of quantum field theory and
is based on action principle and gauge principle. 
Combining general relativity and quantum gauge theory
of gravity, we can determine the cosmological constant
in a simple way, which is well constant with experimental
results. Finally, we will give a possible explanation
on the possible origin of the cosmological constant. \\

\section{Determination of the Cosmological Constant}
\setcounter{equation}{0}    

First, let's discuss the Einstein's field equation
from action principle. The action of the system
is selected as\cite{12}
\be \label{2.1}
A = A_E + A_M,
\ee
\be  \label{2.2}
A_E = \frac{-1 }{16 \pi G} \int {\rm d}^4x
\sqrt{- g} (R - 2 \Lambda),
\ee
\be  \label{2.3}
A_M = \int {\rm d}^4x {\cal L}_M,
\ee
where $G$ is the Newtonian gravitational constant, $R$
is the scalar curvature, $\Lambda$ is the cosmological
constant, ${\cal L}_M$ is the lagrangian density for
matter fields, and 
\be \label{2.4}
g = det (g_{\mu\nu}),
\ee
with $g_{\mu\nu}$ is the space-time metric in general
relativity. Using the following relations\cite{12}
\be \label{2.5}
\delta \sqrt{-g} = \frac{1}{2}
\sqrt{-g } g^{\mu\nu} \delta g_{\mu\nu},
\ee
\be \label{2.6}
\sqrt{-g } g^{\mu\nu} \delta R_{\mu\nu}
= \partial_{\lambda} W^{\lambda},
\ee
\be \label{2.7}
T_m^{\mu\nu} = \frac{2}{\sqrt{-g }}
\frac{\delta A_M}{\delta g_{\mu\nu}(x)},
\ee
where $T^{\mu\nu}$ is the energy-momentum tensor of 
matter fields and $W^{\lambda}$ is a n contravariant 
vector, we can obtain the Einstein's field equation
with cosmological constant $\Lambda$,
\be \label{2.8}
R_{\mu\nu} - \frac{1}{2} g_{\mu\nu} R
+ \Lambda g_{\mu\nu}
= - 8 \pi G T_{\mu\nu},
\ee
where $T_{\mu\nu}$ is the revised energy-momentum tensor,
whose definition is
\be \label{2.9}
T_{\mu\nu} \define T_{m \mu\nu}
- \frac{1}{4 \pi G}
\frac{\delta \Lambda}{\delta g^{\mu\nu}}
\ee
\\

Equation (\ref{2.1}) is one of the starting point of our
following discussions. Another starting point of our
following discussions is from quantum gauge theory of
gravity\cite{11,111}. In literature \cite{11,111}, four different
kinds of action for gravitational field are given. The
best form for our present discussions is
\be  \label{2.9}
A =  \int {\rm d}^4x \sqrt{- g}  {\cal L}_0 + A_M,
\ee
where ${\cal L}_0$ is the lagrangian density for pure
gravitational gauge field whose form is\cite{11,111}
\be  \label{2.10}
{\cal L}_0 = - \frac{1}{4} \eta^{\mu \rho} \eta^{\nu \sigma}
g_{ \alpha \beta}
F_{\mu \nu}^{\alpha} F_{\rho \sigma}^{\beta},
\ee
$F_{\mu \nu}^{\alpha}$ if the field strength of gravitational
gauge field $C_{\mu}^{\alpha}$, which is given by
\be  \label{2.11}
F_{\mu \nu}^{\alpha} = \partial_{\mu} C_{\nu}^{\alpha}
- \partial_{\nu} C_{\mu}^{\alpha}
- g C_{\mu}^{\beta} \partial_{\beta} C_{\nu}^{\alpha}
+ g C_{\nu}^{\beta} \partial_{\beta} C_{\mu}^{\alpha}.
\ee
\\

Suppose that two actions eq.(\ref{2.1}) and eq.(\ref{2.9})
are essential the same, then we will get
\be  \label{2.12}
{\cal L}_0 = \frac{-1}{ 16 \pi G}(R - 2 \Lambda).
\ee
It means the difference of two lagrangians of general relativity
and quantum gauge theory of gravity gives out the
cosmological constant
\be  \label{2.13}
\Lambda = \frac{1}{2} ( R + 4 g^2 {\cal L}_0).
\ee
In eq.(\ref{2.13}), both scalar curvature $R$ and lagrangian
density ${\cal L}_0$ for pure gravitational gauge field
are known, so, we can use eq.(\ref{2.13}) to determine the
cosmological constant $\Lambda$.\\

According to literature \cite{11,111}, both scalar curvature $R$
and lagrangian density ${\cal L}_0$ can be expressed by
gravitational gauge field $C_{\mu}^{\alpha}$. So, the
explicit expression of $\Lambda (x)$ is
\be
\begin{array}{rcl}
\Lambda (x) & = &
2 g^{\mu \kappa} ( \partial_{\mu} G \cdot G^{-1}
\cdot \partial_{\kappa} G \cdot G^{-1} ) ^{\alpha}_{\alpha}
-  g^{\mu \kappa} ( \partial_{\mu} \partial_{\kappa}G
 \cdot G^{-1} ) ^{\alpha}_{\alpha} \\
&&\\                                                                            
&& + \frac{3}{4} \eta^{\rho \sigma } g_{\alpha \beta} g^{\mu \kappa}
( \partial_{\mu} G  ) ^{\alpha}_{\rho}
( \partial_{\kappa} G  ) ^{\beta}_{\sigma}
 -  \eta^{\alpha \beta} G^{\kappa}_{\beta}
( \partial_{\kappa} G \cdot G^{-1} \cdot
\partial_{\lambda} G )^{\lambda}_{\alpha}\\
&&\\                                                                            
&& -  \eta^{\alpha \beta} G^{\kappa}_{\beta}
( \partial_{\lambda} G \cdot G^{-1} \cdot
\partial_{\kappa} G )^{\lambda}_{\alpha}
 +  \eta^{\alpha \beta} G^{\kappa}_{\beta}
( \partial_{\kappa} \partial_{\lambda} G )^{\lambda}_{\alpha}\\
&&\\                                                                            
&& - \frac{5}{4} g_{\alpha \beta} \eta^{\rho \sigma} \eta^{\mu \nu}
G^{\kappa}_{\nu} G^{\lambda}_{\rho}
( \partial_{\lambda} G  ) ^{\alpha}_{\mu}
( \partial_{\kappa} G   ) ^{\beta}_{\sigma} \\
&&\\                                                                            
&& + \frac{1}{2} g_{\alpha \beta} \eta^{\alpha_1 \beta_1} 
\eta^{\mu_1 \nu_1} G^{\nu}_{\beta_1} G^{\mu}_{\mu_1}
( \partial_{\nu} G   ) ^{\alpha}_{\alpha_1}
( \partial_{\mu} G   ) ^{\beta}_{\nu_1 } \\
&&\\                                                                            
&&-  g~ g^{\alpha \beta} F^{\rho}_{\alpha_1 \beta_1}
 G^{-1 \alpha_1}_{\rho}
(G^{-1} \cdot \partial_{\beta} G  \cdot G^{-1} ) ^{\beta_1}_{\alpha} \\
&&\\                                                                            
&& + \frac{1}{2} g \eta^{\alpha \mu} F^{\kappa}_{\mu \beta}
 ( \partial_{\kappa} G  \cdot G^{-1} ) ^{\beta}_{\alpha}
+ \frac{1}{2} g \eta^{\alpha \mu} F^{\rho}_{\alpha \beta}
G^{\kappa}_{\mu}  (G^{-1} \cdot
\partial_{\kappa} G  \cdot G^{-1} ) ^{\beta}_{\rho} \\
&&\\                                                                            
&& + \frac{1}{2} g^2 \eta^{\beta \beta_1} F^{\rho}_{\alpha \beta}
F^{\rho_1}_{\alpha_1 \beta_1} G^{-1 \alpha}_{\rho}
G ^{-1 \alpha_1}_{\rho_1}
- \frac{g^2}{4} \eta^{\alpha \beta } F^{\rho}_{\mu \beta}
F^{\rho_1}_{\alpha_1 \alpha} G^{-1 \alpha_1}_{\rho}
G ^{-1 \mu}_{\rho_1}\\
&&\\                                                                            
&& - \frac{3 g^2}{4}  g_{\alpha \beta} \eta^{\alpha_1 \beta_1}
\eta^{\mu \sigma}  F^{\rho}_{\mu \beta_1}
F^{\beta}_{\sigma \alpha_1 }
\end{array}
\label{2.15}
\ee
From this equation, we can see that cosmological constant is
space-time dependent, so we denoted it as $\Lambda(x)$. The
cosmological constant which affects large scale structure
evolution of cosmos is the average of $\Lambda(x)$ over
the whole Universe.
\\

\section{Estimation of the Cosmological Constant}
\setcounter{equation}{0}

Now, let's make a rough estimation of the cosmological constant.
Suppose that, the moving speed of stars is slow, so
in the vacuum, the dominant component of gravitational
field $C_{\mu}^{\alpha}$ is $C_0^0$ and dominant field strength
of gravitational gauge field is $\svec{E}^0$. Other components
are much smaller than  $C_0^0$ and $\svec{E}^0$, and they
will be neglected in our estimation. In order to estimate
the magnitude of $\Lambda$, we use our solar system as
an example. Denote solar equatorial radius as $r_0$ and
the distance between solar and nearest star as $R_0$.
The gravitational field  $C_0^0$ which is generated by
solar is\cite{11,111}
\be  \label{3.1}
g C_0^0 = - \frac{GM}{r c^2},
\ee 
where $r$ is the distance between a point in space and the sun.
The only non-vanishing field strength $\svec{E}^0$ is
\be  \label{3.2}
g \svec{E}^0 = - \frac{GM}{r^2 c^2} \svec{r}.
\ee
In our estimation, the sun is regarded as quasi-static, so 
all time derivatives vanish. Using the equation of motion 
of pure gravitational gauge field in vacuum, the linear terms
in eq.(\ref{2.15}) vanish\cite{11,111}, 
so we only need to concern quadratic
terms. Using all these relations, 
eq.(\ref{2.15}) gives out the following relation
\be  \label{3.3}
\Lambda (r) = 2 \left (
\frac{GM}{r^2 c^2}  \right )^2
+ o(g C_{\mu}^{\alpha})^3,
~~~~~(r > r_0)
\ee
\be  \label{3.301}
\Lambda (r) = 2 \left (
\frac{GM}{r_0^3 c^2}  \right )^2 \cdot r^2
+ o(g C_{\mu}^{\alpha})^3,
~~~~~(r < r_0).
\ee
The average of $\Lambda(x)$ gives out cosmological constant,
\be  \label{3.4}
\Lambda  = 
\frac{\int_0^{R_0} \Lambda (r) 4 \pi r^2{\rm d}r }
{\int_0^{R_0} 4 \pi r^2{\rm d}r}
\simeq
\frac{36~ G^2 M^2}{5~ R_0^3 r_0 c^4}. 
\ee
The distance between the sun and the nearest star is about
4.5 light year, so
\be  \label{3.5}
R_0 \sim 4.26 \times 10^{16} m.
\ee
The mass of the sun is
\be  \label{3.6}
M = 1.99 \times 10^{30} kg,
\ee
and the radius of the sun is
\be  \label{3.7}
r_0 = 6.96 \times 10^8 m.
\ee
Using all these relation, we get
\be  \label{3.8}
\Lambda \sim 2.92 \times 10^{-52} m^{-2}.
\ee
This value is quite close to experimental results. According
to PDG, the cosmological constant is\cite{13}
\be  \label{3.9}
\Lambda = 3.51 \times 10^{-52} \Omega_{\Lambda} h_0^2 m^{-2},
\ee
with $\Omega_{\Lambda}$ is the scaled cosmological constant 
and $h_0$ is the normalized Hubble expansion rate, whose
values are
\be  \label{3.10}
-1< \Omega_{\Lambda} <2,
\ee
\be  \label{3.11}
0.6 < h_0 < 0.8.
\ee
So, a quite rough estimation of cosmological constant is
quite close to experimental value. \\

\section{Comments}
\setcounter{equation}{0}    

Because $- \left( \frac{GM}{r^2 c^2} \right )^2$  relates
to the energy of gravitational field in vacuum, both
space-time curvature and cosmological constant directly
relate to the average energy of gravitational field in vacuum.
In this paper, a very rough estimation is given. A more 
strict estimation of cosmological constant should average
$\Lambda (x)$ over the vacuum space of the whole Universe.
\\

Because the  energy of gravitational field in vacuum
is negative, it generates a repulsive force on stars
and makes the expansion rate of the Universe accelerated.
This is the reason why cosmological constant is positive. \\

\lbrack {\bf Acknowledgment} \rbrack~~
One of the author (N.Wu) would like to thank 
Prof. Zhe Chang and Prof. Chaoguang Huang for helpful
discussion on cosmological constant. \\

\end{document}